\documentclass[letter]{aa} 

\bibpunct{(}{)}{;}{a}{}{,} 
\usepackage{graphicx}

\usepackage[varg]{txfonts}
\usepackage{makecell}
\usepackage{multirow}
\usepackage{natbib}
\usepackage[english]{babel} 
\usepackage[utf8]{inputenc}
\usepackage[T1]{fontenc}
\usepackage{xspace}
\xspaceaddexceptions{”}

\usepackage{savesym}
\savesymbol{tablenum}
\usepackage{siunitx}
\restoresymbol{SIX}{tablenum}
\usepackage{etoolbox}
\makeatletter
\patchcmd\@combinedblfloats{\box\@outputbox}{\unvbox\@outputbox}{}{}
\makeatother

\begin{document} 

\title{Decayless oscillations in 3D coronal loops excited by a power-law driver}

\newcommand{\orcid}[1]{}
\author{%
{Konstantinos Karampelas}\inst{\ref{aff:CmPA}} \orcid{0000-0001-5507-1891}
\and {Tom Van Doorsselaere}\inst{\ref{aff:CmPA}} \orcid{0000-0001-9628-4113}
}

\institute{%
\label{aff:CmPA}{Centre for mathematical Plasma Astrophysics, Department of Mathematics, KU Leuven, Celestijnenlaan 200B, 3001 Leuven, Belgium.}\\ \email{kostas.karampelas@kuleuven.be}
}

\date{Received 03 October 2023; Accepted 06 December 2023}
 
\abstract
{}
{We studied the manifestation of decayless oscillations in 3D simulations of coronal loops, driven by random motions.}
{Using the PLUTO code, we ran magnetohydrodynamic (MHD) simulations of a straight gravitationally stratified flux tube, with its footpoints embedded in chromospheric plasma. We consider transverse waves drivers with a horizontally polarised red noise power-law spectrum.}
{Our broadband drivers lead to the excitation of standing waves with frequencies equal to the fundamental standing kink mode and its harmonics. These standing kink oscillations have non-decaying amplitudes, and spectra that depend on the  characteristics of the loops, with the latter amplifying the resonant frequencies from the drivers. We thus report for the first time in 3D simulations the manifestation of decayless oscillations from broadband drivers. The spatial and temporal evolution of our oscillation spectra reveals the manifestation of a half harmonic, which exhibits half the frequency of the identified fundamental mode with a similar spatial profile. Our results suggest that this mode is related to the presence of the transition region in our model and could be interpreted as being the equivalent to the fundamental mode of standing sound waves driven on pipes closed at one end.}
{The potential existence of this half harmonic has important implications for coronal seismology, since misinterpreting it for the fundamental mode of the system can lead to false estimations of the average kink speed profile along oscillating loops. Finally, its detection could potentially give us a tool for  distinguishing between different excitation and driving mechanisms of decayless oscillations in observations.}


\titlerunning{Decayless oscillations driven by a power-law driver}
\authorrunning{Karampelas et al.}

\maketitle
\section{Introduction} \label{sec:introduction}

The term decayless transverse (or kink) oscillations has been used over the past decade to describe a category of generally small-amplitude transverse oscillations in coronal loop observations in the extreme ultraviolet (EUV). The name is inspired by the near constant value of their amplitude, persisting over a large number of oscillation periods \citep{nistico2013}. This comes in stark contrast to the other known regime of large amplitude fast decaying transverse loop oscillations \citep{aschwanden1999,nakariakov1999}. Both decaying and decayless transverse oscillations are treated as standing kink oscillations in cylindrical loops \citep[e.g.][]{edwin1983wave,tvd2008detection,anfinogentov2015}, and alongside propagating kink waves they are some of the most studied and ubiquitous phenomena in the solar corona \citep[see][for a review]{NakariakovEtAl2021}. Ever since their discovery in coronal loops \citep{wang2012,tian2012}, the ubiquitous nature of decayless oscillations has been confirmed by a large number of observational studies in coronal loops \citep[e.g.][]{anfinogentov2013,anfinogentov2015,ZhongSihui2022MNRAS.513.1834Z,ZhongSihui2022MNRAS.516.5989Z}, in short coronal loops with length of a tens of Mm \citep[e.g][]{Petrova2023ApJ...946...36P,ShrivastavArpitKumar2023arXiv230413554S}, and in coronal bright points (CBPs) \citep[e.g.][]{GaoYuhang2022ApJ...930...55G}.

Unlike their decaying counterparts, which  are excited by nearby transients \citep[see][]{Nechaeva2019ApJS}, the constant amplitude of these decayless oscillations suggests a continuous supply of energy that will counteract the damping from effects like resonant absorption \citep[e.g.][]{goossens2002coronal}, phase mixing \citep[e.g.][]{heyvaerts1983}, and the Kelvin--Helmholtz instability \citep[KHI;][]{tvd2021ApJ...910...58V,tvd2021ApJ...920..162VErratum}. This makes non-damping kink oscillations a potential heating mechanism for the solar atmosphere (e.g. \citealt{Lim2023ApJ...952L..15L}; see also \citealt{tvd2020coronalSSRv..216..140V} for a review). In addition, decayless oscillations can  be used in coronal seismology, as in \citet{anfinogentov2019ApJ} where they were used to determine the distribution of kink and Alfv\'{e}n speeds in active regions.

However, the excitation mechanism of decayless oscillations is still under debate. In \citet{antolin2016}, they are hypothesised  to be the result of (line of sight) LOS effects due to KHI. \citet{nistico2013} proposed instead a harmonic driver in resonance with the loop, an idea used extensively in simulations \citep[e.g.][]{karampelas2017,mingzhe2019}. However, harmonic drivers lead to equal manifestation of horizontally and vertically polarised oscillations, while observations favour horizontal polarisations \citep{anfinogentov2015}. Excitation via supergranulation flows as a self-oscillation was proposed \citep{nakariakov2016, Nakariakov2022MNRAS.516.5227N}, and was shown numerically \citep{karampelas2020ApJ} to create horizontally linearly polarised oscillations. However, this can not explain decayless oscillations of loops rooted in sunspots, where supergranulation flows are absent \citep{Mandal2022A&A...666L...2M}. Excitation by vortex shedding \citep{nakariakov2009} can also generate non-damping oscillations \citep{karampelas2021ApJ...908L...7K} and could explain observations of flare-induced decayless  oscillations in \citet{Mandal2022A&A...666L...2M}. \citet{afanasyev2020decayless} explored the spectrum of excited oscillations from a red noise driver in a 1D study, providing the first proof of concept for broadband drivers exciting multiple harmonics in decayless oscillations \citep[see also][]{ruderman2021MNRAS.501.3017R,ruderman2021SoPh..296..124R}. The contribution of footpoint-driven transverse waves with a multi-frequency spectrum in the energy evolution and heating of 3D coronal loops has also been studied in recent years \citep[e.g.][]{Pagano2019,Pagano2020A&A...643A..73P,Howson2023Physi...5..140H}. However, no numerical study of 3D coronal loops perturbed by broadband drivers has reported on the excitation and evolution, spatial and temporal, of standing waves matching the observed decayless oscillations, in the presence or absence of a transition region and a lower chromosphere.

In this letter we study the spectra of decayless oscillations in a 3D gravitationally stratified loop, generated by a driver with a power-law spectrum. In Section \ref{sec:setup} we  describe our initial and boundary conditions and the numerical scheme used throughout our simulations. Our findings are presented in detail in Section \ref{sec:results}. Finally, a thorough discussion of our results and their implications for the topic focused on here takes place in Section \ref{sec:discussions}.

\section{Numerical set-up} \label{sec:setup}

\begin{figure}[t]
    \centering
    \includegraphics[trim={0.cm .4cm 0.cm 0.cm},clip,scale=0.55]{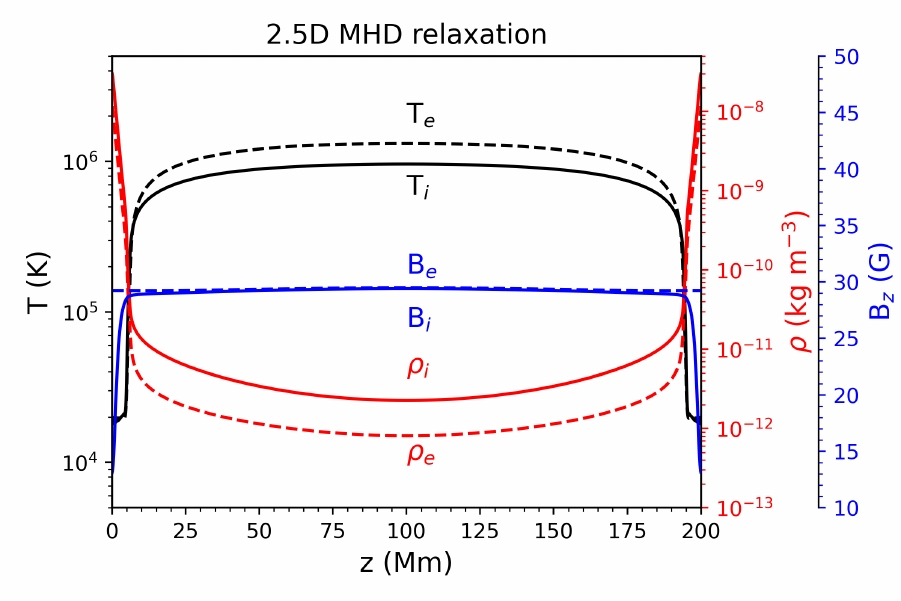}
    \caption{Temperature, density, and $B_\mathrm{z}$ magnetic field along $z$ inside (solid line) and outside (dashed line) of the flux tube, at the end of the 2.5D MHD relaxation.}    \label{fig:hydrostatic}
\end{figure}

\begin{figure*}[t]
    \centering
    \includegraphics[trim={0.cm 0.4cm 0.cm 0.cm},clip,scale=0.45]{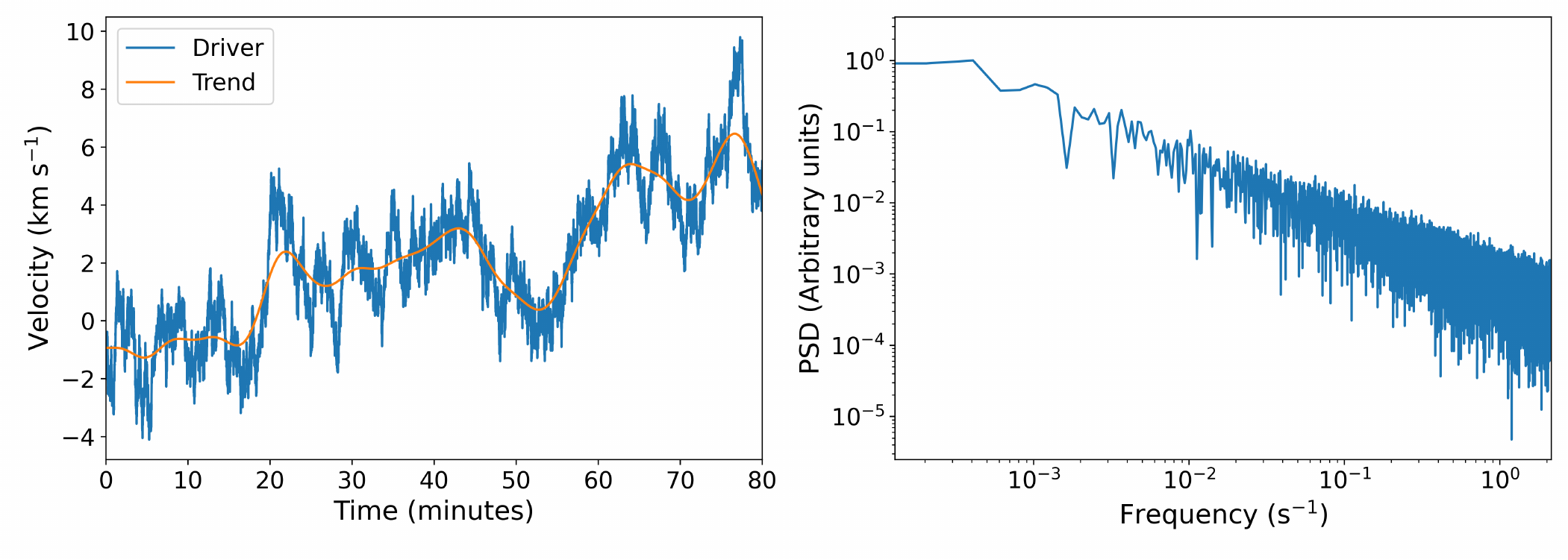}
    \caption{Velocity signal (left panel) and spectrum (right panel) for our red noise driver. The orange line shows its velocity trend.}  \label{fig:Vprofile}
\end{figure*}

\begin{figure*}[t]
    \centering
    \includegraphics[trim={0.cm .8cm 0.cm 0.cm},clip,scale=0.45]{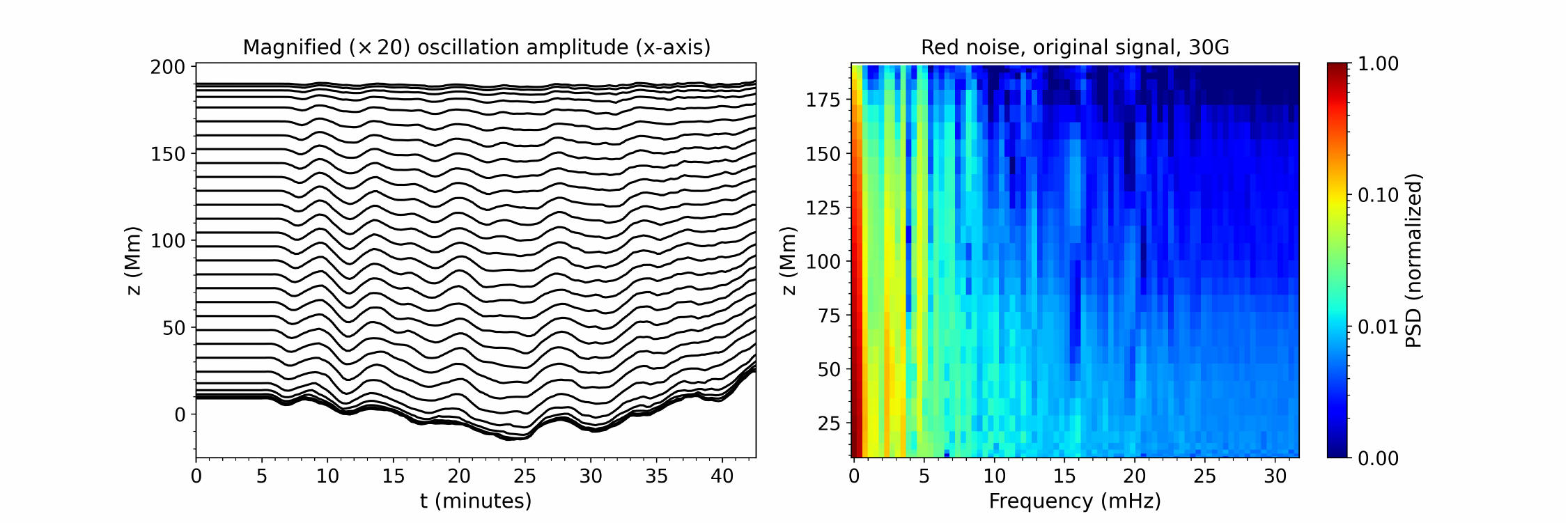}
    \includegraphics[trim={0.cm .8cm 0.cm .8cm},clip,scale=0.45]{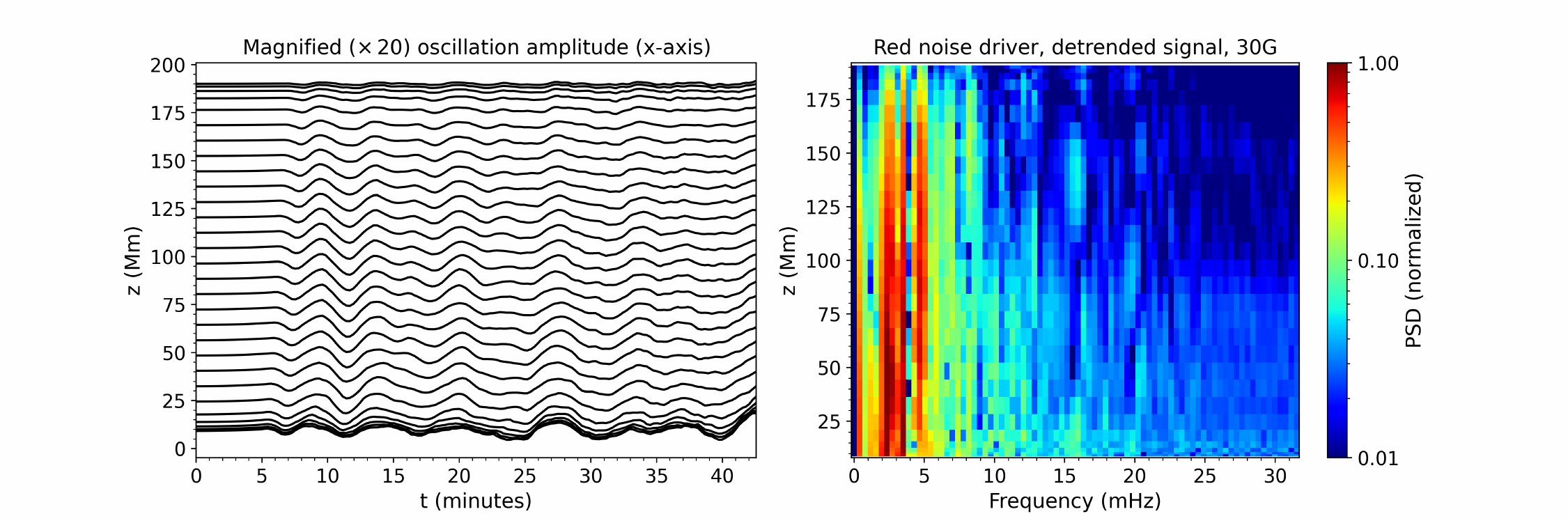}
    \includegraphics[trim={0.cm 0.cm 0.cm .8cm},clip,scale=0.45]{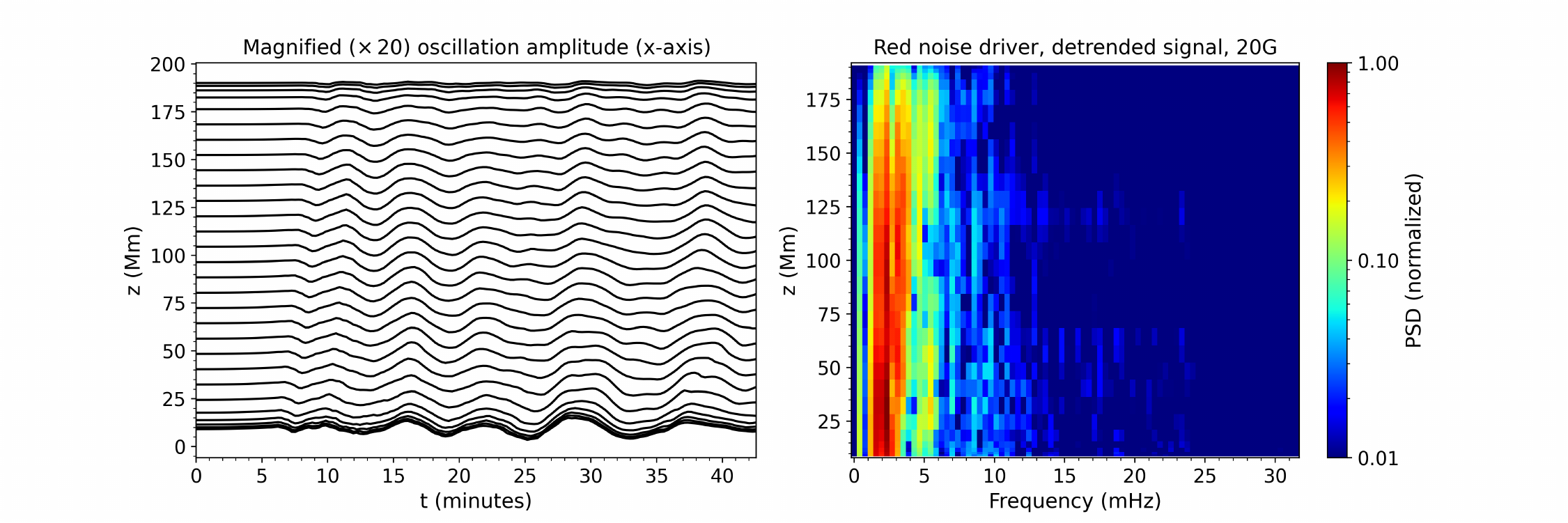}
    \caption{Centre of mass $x$ displacement and corresponding spectra for two different loops. Left panels: Magnified ($\times 20$) $x$ displacement of the loop centre of mass per height (projected on $z$) from slits placed along the coronal part of the loop. Right panels: Fourier power spectra of the displacement taken from the slits on the left. The top and middle panels correspond to the original and detrended displacement signals for a loop with a coronal magnetic field $B_\mathrm{z}\sim 30$\,G. The bottom panels correspond to the detrended displacement signal for a loop with $B_\mathrm{z}\sim 20$\,G.}   \label{fig:fourierspectra}
\end{figure*}

\begin{figure*}[t]
    \centering
    \includegraphics[trim={0.2cm 0.2cm 1.cm 0.2cm},clip,scale=0.48]{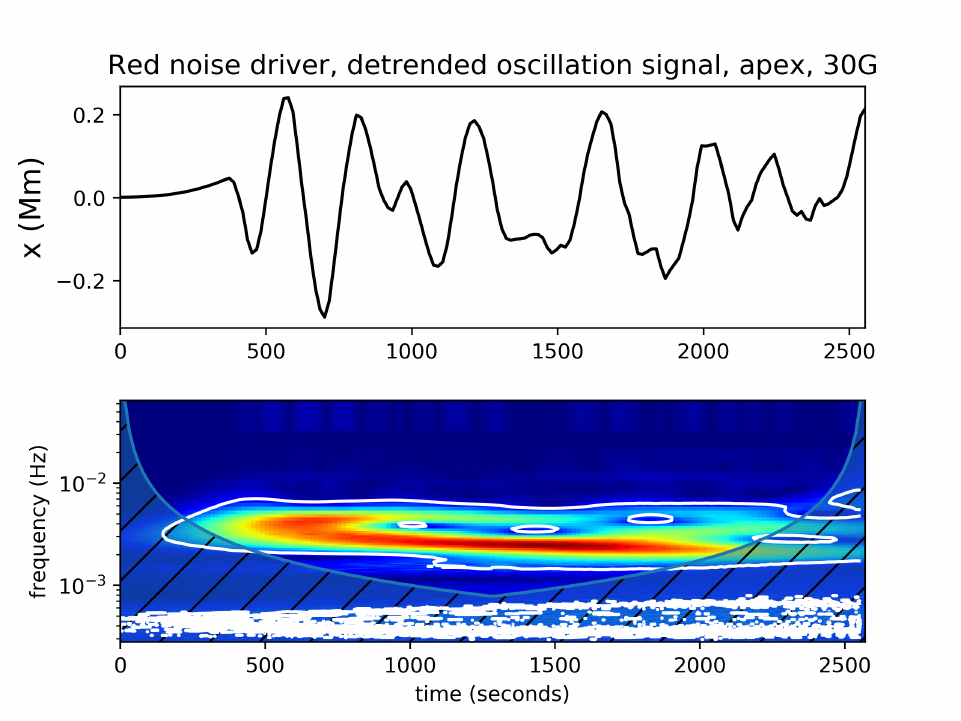}
    \includegraphics[trim={0.2cm 0.2cm 1.cm 0.2cm},clip,scale=0.48]{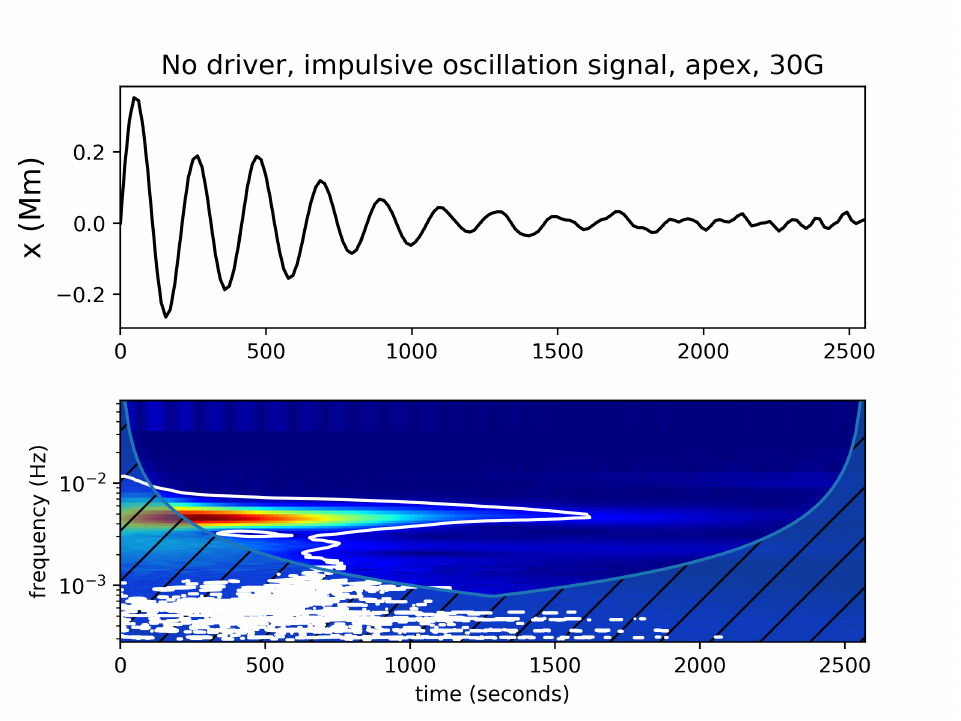}
    \includegraphics[trim={0.2cm 0.2cm 1.cm 0.2cm},clip,scale=0.48]{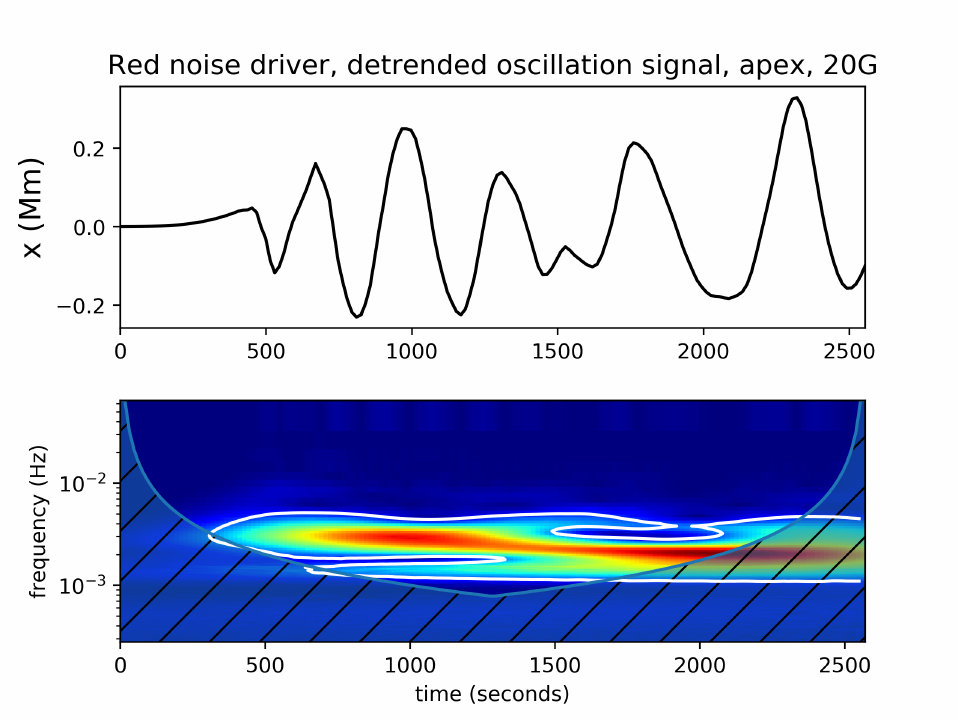}
    \includegraphics[trim={0.2cm 0.2cm 1.cm 0.2cm},clip,scale=0.48]{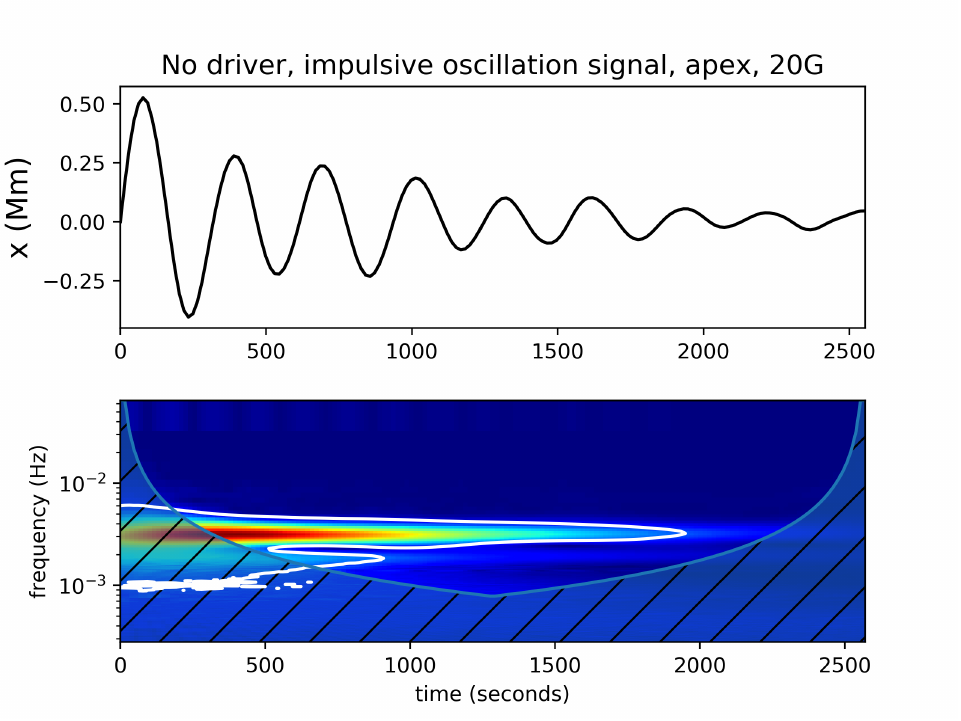}
    \caption{Displacement of the loop apex ($z=100$\,Mm) and respective wavelet spectra, for a loop with $B_\mathrm{z}\sim 30$\,G (top panels) and $B_\mathrm{z}\sim 20$\,G (bottom panels). The signals on the left are the detrended displacement from the driven oscillations. The signals on the right correspond to the fundamental standing kink mode, excited by an initial velocity field. The white contours show the $99\%$ significance level.} 
    \label{fig:30Gwavelets}
\end{figure*}

\begin{figure*}[t]
    \centering
    \includegraphics[trim={0.cm 0.cm 0.cm 0.cm},clip,scale=0.41]{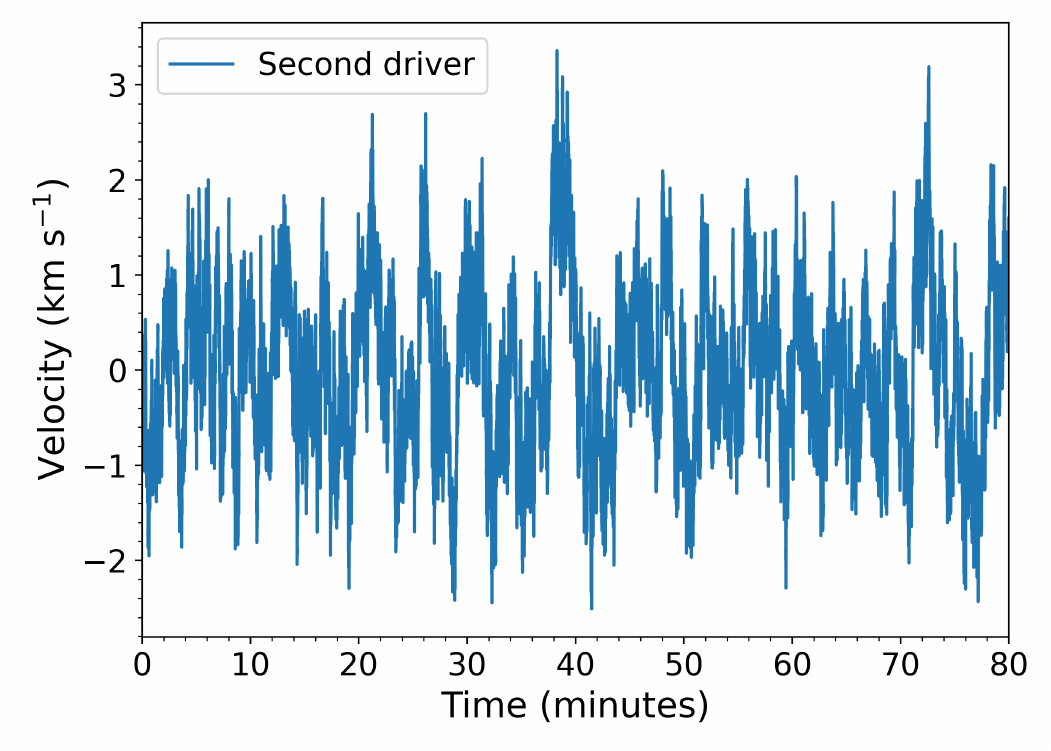}
    \includegraphics[trim={0.cm 0.cm 0.cm 0.cm},clip,scale=0.45]{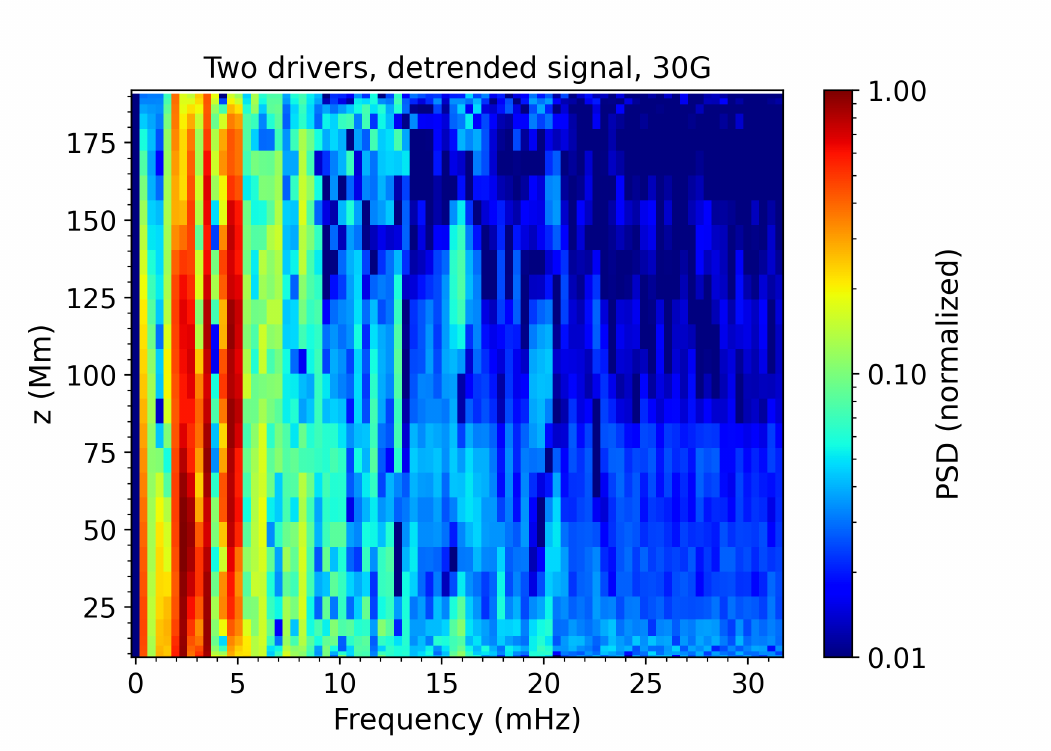}
    \caption{Signal of the second velocity driver at $z=200$\,Mm (left panel) and PSD of the respective centre of mass displacement (right panel) for a loop with a coronal magnetic field $B_\mathrm{z}\sim 30$\,G, driven from both footpoints. The original driver is also active at $z=0$.}   \label{fig:2drivers}
\end{figure*}

\textit{Initial conditions:} We modelled a coronal loop with a chromospheric part and a transition region, following  \citet{pelouze2023A&A...672A.105P} and \citet{mingzhe2023ApJ...949L...1G}. We calculated the hydrostatic equilibrium in a 2.5D slice with $r\in [0,8]$\,Mm and $z\in [0,200]$\,Mm, a uniform grid ($200\times 2048$ points), and sinusoidal gravity ($g_z=274\,\cos(\pi\,z/200)$\,m s$^{-2}$) along the vertical magnetic field lines $\mathbf{B}= B_\mathrm{z} \hat{k}= 30$\,G, using the following profiles: 
\begin{eqnarray}  
\small T(r,z) = \left \{
    \begin{array}{ll}
    T_\mathrm{Cor}, & \text{for $5\text{ Mm} \leq z \leq 195\text{ Mm}$}, \\
    T_\mathrm{Ch}, & \text{elsewhere,} 
    \end{array}\right.  \\
\small T_\mathrm{Cor} = T_\mathrm{Ch} + (T_\mathrm{C}(r) - T_\mathrm{Ch})( 1 - \left[ (200-z)/195 \right]^2 )^{0.3},  \\
T_\mathrm{C}(r) = T_\mathrm{C,e} + (T_\mathrm{C,i} - T_\mathrm{C,e})\, \zeta(r),   \\
\rho_\mathrm{Ch}(r) = \rho_\mathrm{Ch,e} + (\rho_\mathrm{Ch,i}-\rho_\mathrm{Ch,e})\, \zeta(r),  \\
\zeta(r) = 0.5 [ 1 - \tanh\left(\left( [r/R]-1 \right) 20 \right) ]. 
\end{eqnarray}
The subscripts $\mathrm{Ch}$, $\mathrm{Cor,}$ and $\mathrm{C}$ refer to the chromospheric ($z=0$ and $z=200$\,Mm), coronal, and apex ($z=100$\,Mm) values, respectively, inside ($i$) and outside ($e$) of the loop. The function $\zeta(r)$ gives us the radial loop profile, with a cross-section radius of $R=1$\,Mm. Finally, we have $T_\mathrm{Ch}=0.02$\,MK, $T_\mathrm{C,i} = 1$\,MK, $T_\mathrm{C,e} = 1.5$\,MK, $\rho_\mathrm{Ch,i} = 3.51 \times 10^{-8}$\,kg m$^{-3}$, and $\rho_\mathrm{Ch,e} = 1.17 \times 10^{-8}$\,kg m$^{-3}$. We let the system evolve for $3890$\,s, reducing each velocity component per iteration ($v_i = v_i/1.0001$) for $2334$\,s, reaching a semi-equilibrium state with maximum residual velocity values of $\sim 4$\,km s$^{-1}$ along the z direction and  $\sim 0.05$\,km s$^{-1}$ along the x and y directions. The initial conditions after the 2.5D MHD relaxation are shown in  Figure \ref{fig:hydrostatic}. We then interpolated the results on a 3D Cartesian grid with dimensions $x\in [-6,6]$\,Mm, $y\in [0,4]$\,Mm, and $z\in [0,200]$\,Mm. We considered $\delta x = \delta y = 40$\,km everywhere, $\delta z = 98$\,km for $z\leq 10$\,Mm and $z\geq 190$\,Mm, $\delta z = 800$\,km for $z\in [24, 176]$\,Mm, and a stretched grid for $z\in (10, 24)$\,Mm and $z\in (176, 190)$\,Mm. 

\textit{Boundary conditions:} In the 2.5D set-up, we consider axisymmetry at $r=0$ and open boundaries at $r=8$\,Mm. At the $z=0$ and $z=200$\,Mm, we take zero-gradient conditions for the magnetic field, antisymmetry for the velocity, symmetry for the density, and constant temperature $T_\mathrm{Ch}$. For the 3D set-up, we have open boundaries at $x=-6$\,Mm, $x=6$\,Mm, and $y=4$\,Mm and a reflective boundary at $y=0$, simulating half the loop. At $z=0$ and $z=200$\,Mm, the boundaries are the same as in the 2.5D case, but with added symmetry for the pressure. At $t_0 = 202$\,s, we apply a velocity driver ($\lbrace v_x, v_y \rbrace = \lbrace V(t)\,\zeta(r,t), 0\rbrace$) at $z=0$, where we take $R=R_d=2.5$\,Mm in $\zeta(r,t)$, to ensure that the loop always remains inside the area of the driver. For the velocity signal $V(t)$, we take a red noise power-law  spectrum ($S\propto f^{-1.66}$, with $f$ being the frequency), using the colorednoise 2.1.0 Python package. Our driver is shown in Figure \ref{fig:Vprofile};  the orange line depicts its background trend, calculated with a low pass filter. Our driver also  tracks the location of the footpoint ($r(t)$) by numerically integrating the velocity signal over time.

\textit{Numerical scheme:} We solve the full magnetohydrodynamic (MHD) equations for a hydrogen plasma with the PLUTO code \citep{mignonePLUTO2007}, using the parabolic method in 2.5D and the MP5 method in 3D, the Roe solver, the third-order Runge--Kutta method, and the extended GLM formulation. The effective numerical diffusivity ($\eta$) is estimated at $\sim 10^{-5} - 10^{-4}$ in units of the inverse magnetic Reynolds number. In the 2.5D set-up, we also add explicit magnetic diffusivity $\eta = 10^{-4}$. We include thermal conduction  ($\kappa_{\parallel} = 9 \times 10^{-12} T^{5/2}$ in J\,s$^{-1}$\,K$^{-1}$\,m$^{-1}$) and saturation effects for large temperature gradients. For $T\leq T_\mathrm{cut} = 0.25$\,MK, we apply the correction by \citet{LinkerEtAl2001}, \citet{LionelloEtAl2009}, and \citet{MikicEtAl2013}, to treat the transition region with a coarser grid.

\section{Results} \label{sec:results}

We limit our analysis to the coronal part of the loop, while using the lower chromospheric part as a mass reservoir. As is     shown below, our driver displaces the loop while also exciting kink oscillations. To study these oscillations, we track the horizontal position of the loop centre of mass for each $z$ plane, by calculating the average displacement, using as weight the quantity $(\rho(z)-\rho_e(z))^2$, where $\rho_e(z)$ is the density outside the loop. The averaging takes place across the entire domain at each z-plane. This gives us the transverse displacement of the loop along $x$ over time, for each height. 

The top left panel of Figure \ref{fig:fourierspectra} shows the displacement of the coronal part of the loop for waves excited by the red noise driver. We  considered a number of `slits' ($21$ for $z\in\left[25,175\right]$) and projected the signal along $z$. For better visualisation we also magnified the signal of each slit by simply multiplying by a factor of $20$. This multiplication is not employed anywhere else in our analysis, but is only used to visually show the loop motion. The right panel shows the normalised Fourier power spectral density (PSD), along the loop, with resolution along $z$ corresponding to the number of slits used. The low frequency components of the red noise driver lead to large a displacement of the driven footpoint, diminishing as we travel along $z$ towards the anchored footpoint. This aperiodic displacement also manifests in the spectra shown in the top right panel of Figure \ref{fig:fourierspectra}, for frequencies close to $0$. We also see the spatial profile of the corresponding frequencies, being stronger closer to the driven footpoint. 

In order to focus on the periodicities of the observed oscillating patterns in the top left panel of Figure \ref{fig:fourierspectra}, we removed the background trend of the aperiodic displacement from our signal by applying a high pass Gaussian filter, similarly to what could have been applied on a signal derived from EUV observations of decayless oscillations. This process is also the same as that   used to calculate the background trend of our driver, shown in Figure \ref{fig:Vprofile}. The now detrended displacement and respective Fourier spectra are shown in the middle panels of Figure \ref{fig:fourierspectra}. In the middle right panel, we now clearly see two bands of frequencies, one centred at $\sim 2.5-3$\,mHz and one at $\sim 4-5$\,mHz. The nature of these bands, which  are also visible in the spectrum of the full signal shown in the top right panel of the same figure, can be understood by identifying the fundamental kink mode of our loop. 

The top left panels of Figure \ref{fig:30Gwavelets} show the non-magnified detrended displacement signal at the apex and its corresponding wavelet spectrum over time, for our loop. We see that the detrended signal depicts a transverse oscillation with amplitudes of the order of $\sim 0.2$\,Mm. This amplitude shows no significant decay over time, and is typical of the values found in decayless oscillations of coronal loops \citep[e.g.][]{anfinogentov2015}. The two frequency bands at $\sim 2.5-3$\,mHz and $\sim 4-5$\,mHz can also be seen here, persisting over time, with the former exhibiting an overall stronger signal. As a control, we  also performed a simulation of the same loop oscillating with a non-driven decaying fundamental standing kink mode, excited by an initial velocity field. The signal and wavelet spectrum is shown in the top right panels of Figure \ref{fig:30Gwavelets}, from which we confirm that the $\sim 4-5$\,mHz frequency band corresponds to the fundamental standing kink mode for our oscillating loop. Looking again at the middle right panel of Figure \ref{fig:fourierspectra}, we see that the spatial profile of this frequency band indeed matches the expected one for the fundamental standing kink mode. Having oscillations with the period of the fundamental standing kink mode and a non-decaying amplitude, we are led to describe these excited standing waves as decayless oscillations.

Alongside the fundamental mode, in the middle right panel of Figure \ref{fig:fourierspectra} we can also see patterns of additional higher harmonics: the second ($\sim 7.5-8$\,mHz), third ($\sim 12$\,mHz), fourth ($\sim 16$\,mHz), and even fifth ($\sim 20$\,mHz) harmonic. These were made clear after the removal of the contribution from the lower frequencies. However, they can also be detected in the PSD contours in the top right panel of  Figure \ref{fig:fourierspectra}, although their signal is much weaker than that of the low frequency loop displacement of the original time series.

The $\sim 2.5-3$\,mHz frequency band, which we descriptively refer to as the `half harmonic', has frequencies around half that of the identified fundamental mode. This frequency band is also present within the $99\%$ significance level contours in the wavelet spectra of the decaying fundamental kink mode (see top right panels of Figure \ref{fig:30Gwavelets}). As can be seen by the wavelets, this mode is only weakly excited by the initial velocity field. Finally, both the half harmonic  and the fundamental frequency bands have finite widths in the PSD graph, due to the limited sample size of the signal, as well as the development of instabilities and the restructuring of the loop density. 

To ensure that the identified frequency bands are the result of the loop filtering and magnifying its natural frequencies, and not exclusively the result of the power distribution of our driver, we  performed an additional set of simulations, one with the red noise driver and one with an impulsive oscillation, for a different loop, with initial coronal magnetic field of $B_\mathrm{z}=20$\,G. This loop has  temperature and density profiles very similar to those shown in Figure \ref{fig:hydrostatic} for the loop with $B_\mathrm{z}=30$\,G. The detrended displacement and corresponding Fourier spectrum is shown in the bottom panels of Figure \ref{fig:fourierspectra}. The bottom panels of Figure \ref{fig:30Gwavelets} again show the signal and wavelet spectrum at the apex for the detrended displacement and for an impulsive decaying oscillation. By studying the Fourier and wavelet PSDs, we see that the new fundamental mode ($\sim 3$\,mHz) and its corresponding half harmonic ($\sim 1.5$\,mHz) and overtones (second at $\sim 6$\,mHz, third at $\sim 9$\,mHz) are excited.

Finally, we  performed a simulation of our original loop with a coronal magnetic field of $B_z=30$\,G, where we simultaneously employ two drivers: (a) the original one (see Figure \ref{fig:Vprofile}) at $z=0$ and (b) a new driver at $z=200$\,Mm. The second driver, shown in the left panel of Figure \ref{fig:2drivers}, was created from a new red noise spectrum, different from that of the original driver, with the additional step of removing the low frequency background trend. Driving the loop from both footpoints led again to the superposition of a decayless oscillation and a low frequency displacement. Taking the PSD of the detrended displacement signal in the right panel of Figure \ref{fig:2drivers}, we see its similarity to that in the middle right panel of Figure \ref{fig:fourierspectra}. The  main difference comes from the second driver, which  increases the power of the half harmonic near the second footpoint.

\section{Discussion and conclusions} \label{sec:discussions}

In this study we report for the first time in 3D MHD simulations the excitation of decayless kink oscillations driven by a footpoint power-law driver, for a gravitationally stratified coronal loop with footpoints embedded in chromospheric plasma. The amplitudes of our oscillations are of the order of $0.2$\,Mm and do not show any signs of decay for the duration of our simulations ($\sim 5-7$ oscillation periods, depending on the characteristics of each loop used). Figures \ref{fig:fourierspectra} and \ref{fig:30Gwavelets} also show a modulation of the oscillation amplitudes, similar to what was reported in \citet{Nakariakov2022MNRAS.516.5227N} for a model of self-oscillations with additional random-motion terms. However, further exploration of this effect with respect to past studies of self-oscillations is required, which falls outside the scope of this present work. The low frequency component in our driver does not suppress the decayless oscillations when the fully compressible 3D MHD equations are considered, but instead adds an overlaying displacement that appears aperiodic. Expanding upon this, driving a loop at any frequency band would manifest in the oscillation spectrum, for example  in simulations of decayless oscillations of short coronal loops driven by p-modes \citep{GaoYuhang2023ApJ...955...73G}. Using a high pass Gaussian filter to remove this aperiodic, very low frequency motion, we again end up with a signal of a decayless oscillation.  

Using loops of different initial conditions, we see that the same driver leads to the excitation of standing modes for the resonant frequencies corresponding to each loop. This further supports the findings of the 1D study in \citet{afanasyev2020decayless}, proving that coronal loops can filter the driving frequencies in broadband drivers by responding to the resonant ones. We also see our loops manifesting the fundamental standing kink mode and its overtones and we identify modes as high as the fifth harmonic. Such overtones, mainly the second and third harmonic, have also been observed in kink oscillations of loops in EUV, both for decayless \citep[e.g.][]{duckenfield2018ApJ} and decaying \citep[e.g.][]{duckenfield2019A&A...632A..64D} oscillations, with our model also reproducing them for random motion footpoint driving in the case of decayless oscillations. 

In addition to the fundamental mode and its overtones, our model also predicts the excitation of a half harmonic for each loop, with wavelength $\lambda_\mathrm{half}=4L$, where $L$ is the loop length. This frequency band scales with the fundamental, with faint traces of it detected in the spectra of decaying kink oscillations. Its scaling with the fundamental mode is difficult to explain, as it does not agree with the expected solutions of the wave equation in a stratified medium with a transition region present \citep[see][for the harmonics in a 1D model using Alfv\'{e}n waves]{HowsonBreu2023MNRAS.526..499H}. A possible interpretation can be given by considering the effects of the transition region, which is not a perfectly reflective boundary \citep{pelouze2023A&A...672A.105P}. The wave reflections there will set up standing waves, establishing the fundamental standing kink modes and its overtones that we detect. At the same time, the movement of the footpoints in the transition region will create a velocity antinode there. As the loops oscillate, this will lead to a superposition of waves, each seeing one `closed' and one `open' footpoint, leading to the manifestation of the half harmonic with a frequency that is half that of the fundamental. The equivalent mechanical analogue is that of the fundamental mode of standing sound waves in cylindrical pipes closed at only one end (like pan flutes) having half the frequency of the fundamental in open-ended (or closed-ended) pipes. The lower velocities in the transition region for footpoints anchored at the chromosphere, could explain why our simulations of decaying oscillations also show faint traces of this half harmonic.

The exact nature of this half harmonic needs to be further explored in a dedicated study. However, this predicted half harmonic can have profound effects in coronal seismology. The ratio of the periods from multiple harmonics detected in standing loop oscillations can be used to calculate the profile of the average kink speed ($C_K$) along the loop \citep{JainHindman2012A&A...545A.138J}. For example, the models by \citet{andries2005} and \citet{Safari2007} attribute deviations from the expected values of these ratios to density stratification, calculating the periods of the modified first, second, and third harmonics as
\begin{eqnarray}
    P_1 = P_{kink}(1+L/(3\pi^2H))^{-1}, \quad\\
    P_2 = P_{kink}(2+2L/(15\pi^2H))^{-1},\\
    P_3 = P_{kink}(3+3L/(35\pi^2H))^{-1}.
\end{eqnarray}
Here $P_{kink}=2L/C_K$ is the period of the fundamental standing kink mode and $H$ the density scale height. A half harmonic with a similar spatial profile to the first harmonic, but with significantly more power, as our Fourier analysis suggests, could be misidentified as the $P_1$ mode, leading to false estimations of the density scale height for the model previously mentioned, the magnetic field or the $C_K$ profile along the loop in general. Misidentifying this half harmonic could also lead to errors in the slopes of a factor of $2$, when studying the correlation of the oscillation period and the loop length \citep[e.g.][]{anfinogentov2015,ShrivastavArpitKumar2023arXiv230413554S}. Therefore, this predicted half harmonic, if observed, can be of great importance when using decayless oscillations as tools for coronal seismology.

Finally, we note that such a half harmonic would be more prominent if the loop is driven by displacing its footpoints with such broadband drivers. Models of decayless oscillations that require both footpoints to be anchored or that they move primarily with very small amplitudes at dissonant frequencies, will exhibit fainter traces of this mode. It remains to be tested whether the detection of the half harmonic, or lack thereof, could give us a tool to distinguish between footpoint driving and alternative excitation mechanisms for decayless oscillations in loops, such as driving by vortex shedding, or generating self-oscillations through the loop interacting with supergranulation flows.

\begin{acknowledgements}
We would like to thank the referee whose review helped us improve the manuscript. K.K. acknowledges support by an FWO (Fonds voor Wetenschappelijk Onderzoek – Vlaanderen) postdoctoral fellowship (1273221N). TVD was supported by the European Research Council (ERC) under the European Union's Horizon 2020 research and innovation programme (grant agreement No 724326), the C1 grant TRACEspace of Internal Funds KU Leuven, and a Senior Research Project (G088021N) of the FWO Vlaanderen.
\end{acknowledgements}

\bibliographystyle{aa}
\bibliography{paper}

\end{document}